\documentclass[11pt]{article}
\usepackage{geometry}                
\usepackage{graphicx}
\usepackage{epsfig}
\usepackage{amssymb}
\usepackage{epstopdf}
\title{Coherent $\rho$ and $J/\psi$ photoproduction in ultraperipheral processes 
with electromagnetic dissociation of heavy ions at RHIC and LHC.}
\author {
V.~Rebyakova\\
\it Petersburg State Technical University\\
\it  St.Petersburg, Russia\\
M. ~Strikman\\
\it Pennsylvania State University,\\
\it University Park, Pennsylvania  16802, USA
\\
M.~Zhalov\\
\it Petersburg Nuclear Physics Institute, Gatchina, 
Russia}

\begin{document}
\maketitle
\section*{Abstract}

We present predictions for the $J/\psi$ and $\rho$ meson production in 
the heavy ion ultraperipheral collisions (UPC) for the current energy 2.76 TeV at the LHC.
Both total cross sections and cross sections with the neutron emission from one or both nuclei are presented.   
We also perform analysis of the RHIC $\rho$-meson photoproduction data and emphasize importance of 
these data to test the current model for nucleus break up in UPC.
\section{Introduction}
Ultraperipheral collisions at the LHC provide a  fine probe for studies of the 
small x dynamics in much cleaner environment than strong proton - proton (nucleus) collisions. 
They also provide an important bridge to the future studies at the e-A colliders.
A detailed study of the UPC physics was performed by the UPC study group and reported in \cite{Baltz:2007kq}. 
However, all these studies were performed for the planned energy of the heavy ion collisions 
at the LHC - $\sqrt{s}\mbox{= 5.52 TeV}$. Hence, it is necessary to provide the predictions for the 
current energy $\sqrt{s}\mbox{= 2.76 TeV}$ in models, used in \cite{Baltz:2007kq},
as well as to explore long term advantages of having UPC data at two different LHC energies.
 Also, we receive requests from  experimental groups to calculate the quantities  which are more  
readily accessible experimentally - the partial cross sections with dissociation of one or 
two nuclei with emission of a number of neutrons.

The basic expressions for the cross section of production of vector mesons in ultraperipheral 
collisions are given in a number of papers, see e.g. \cite{Baltz:2007kq}:
\begin{eqnarray}
{\frac {d\sigma_{A_{1}A_{2}\to A_{1}A_{2}V}} {dy}}=
n_{\gamma /A_{1}}(y)\sigma_{\gamma A_{2}\to V A_{2}}(y)+
n_{\gamma /A_{2}}(-y)\sigma_{\gamma A_{1}\to V A_{1}}(-y).
\end{eqnarray}
The quantity $y=ln \frac {2\omega_{\gamma } } {M_{V}}$ is the rapidity 
of the produced vector meson, and $n_{\gamma /A}(y)$ is the flux of photons 
with the energy $\omega _{\gamma }=\gamma _{L}q_0$ 
emitted by one of nuclei ($\gamma _{L}$ is 
the Lorentz factor for colliding nuclei, and
$q_{0}$ -is the photon momentum in the coordinate system of moving nucleus).

The specific feature of the ultraperipheral collisions is rather large impact parameter $\vec b$
between interacting nuclei. Hence, the flux of photons, radiated by accelerated ion, can be with
reasonable accuracy approximated by the equivalent photon spectrum of the point charge $Z$, moving
with velocity $\beta=p_N /E_N$.
To avoid the strong interaction of the colliding ions 
(which breaks coherence of the process),
 one  has to  restrict the impact parameters $b$ of the interaction by 
choosing for example  $b_{min} \approx 2 R_A$ or 
by  suppressing the strong interactions by the factor 
$P_{S} (b)=\exp \bigl [-\sigma_{NN}T_{AA}(b)\bigr ]$ 
with
$$T_{AA}(b)=\int \limits_0^\infty db^\prime \int \limits_{-\infty}^{\infty} dz 
\rho_{A}({\vec b}-{\vec b^\prime},z)\int \limits_{-\infty}^{\infty}
\rho_A ({\vec b}^{\prime},z^{\prime} )dz^\prime
$$ 
 describing overlap of colliding ions ($\rho_A ({\vec b},z)$ is the nuclear density
normalized by the condition $\int d^2bdz \rho ({\vec b},z)=A$).
However,  due to the large value of $Z$ in UPC of heavy nuclei, there is a significant 
probability of the photon exchange  with excitation of nuclei accompanied by a subsequent decay 
with emission of neutrons \cite{Baltz}.
When such additional photon exchange occurs in the final state of UPC 
with the vector meson photoproduction, this does not destroy the coherence of photoproduction, 
and sum over the final state of ions  gives 
  the total coherent photoproduction cross section. On the other hand, measurements by STAR 
at RHIC demonstrate that, using the Zero Degree
Calorimeters (ZDC) which are installed in all detectors at RHIC and LHC,
 it is possible to select different 
channels, in particular, coherent photoproduction with excitation of ions in which 
 neutrons are emitted by one or both nuclei. One can distingish several channels:  1n1n - emission of 
one neutron by each ion, XnXn - emission of several neutrons, 0n1n and 0nXn  excitation and decay 
only of one ion.
To estimate these partial cross sections,
we use the model suggested in \cite{Baltz} where it is shown that 
coherent photoproduction with additional electromagnetic nucleus excitation 
can be calculated modifying the flux of photons 
by impact parameter dependent factors $P_{C}^{i} ({\vec b})$ which account for different channels  
(i=0n0n, 0n1n,1n1n,0nXn, XnXn)
\footnote{Our thanks  to S.Klein and J.Nystrand for the kindly providing  
their code for calculating
the probabilities of electromagnetic excitation with subsequent neutron decay}. 
Hence, the final expression for the
photon flux used in our calculations is

\begin{equation}
n_{\gamma /A}^{i}(\omega_{\gamma} )= \frac {2 \alpha Z^2} {\pi} \int \limits_{b_{min}}^{\infty} db\frac {x^2} {b} \biggl [K_{1}^2 (x)
+\frac {K_0^2 (x)} {\gamma_{L}^2}\biggr ]P_S ({b})P_{C}^{i}(b),
\label{fluxf}
\end{equation}
where $x=\frac {\omega_{\gamma} b} {\gamma_L \beta}$, and $K_0 , K_1$
are the modified Bessel functions. 

We provide predictions for the light and heavy vector meson photoproduction with small transverse 
momenta of vector mesons.
The dynamics of production of mesons, built of light quarks, and those, built of heavy quarks, 
is qualitatively different. 
Hence, we treat them separately.

\section{Production of light vector mesons}

\begin{figure}
\begin{center}
\epsfig{file=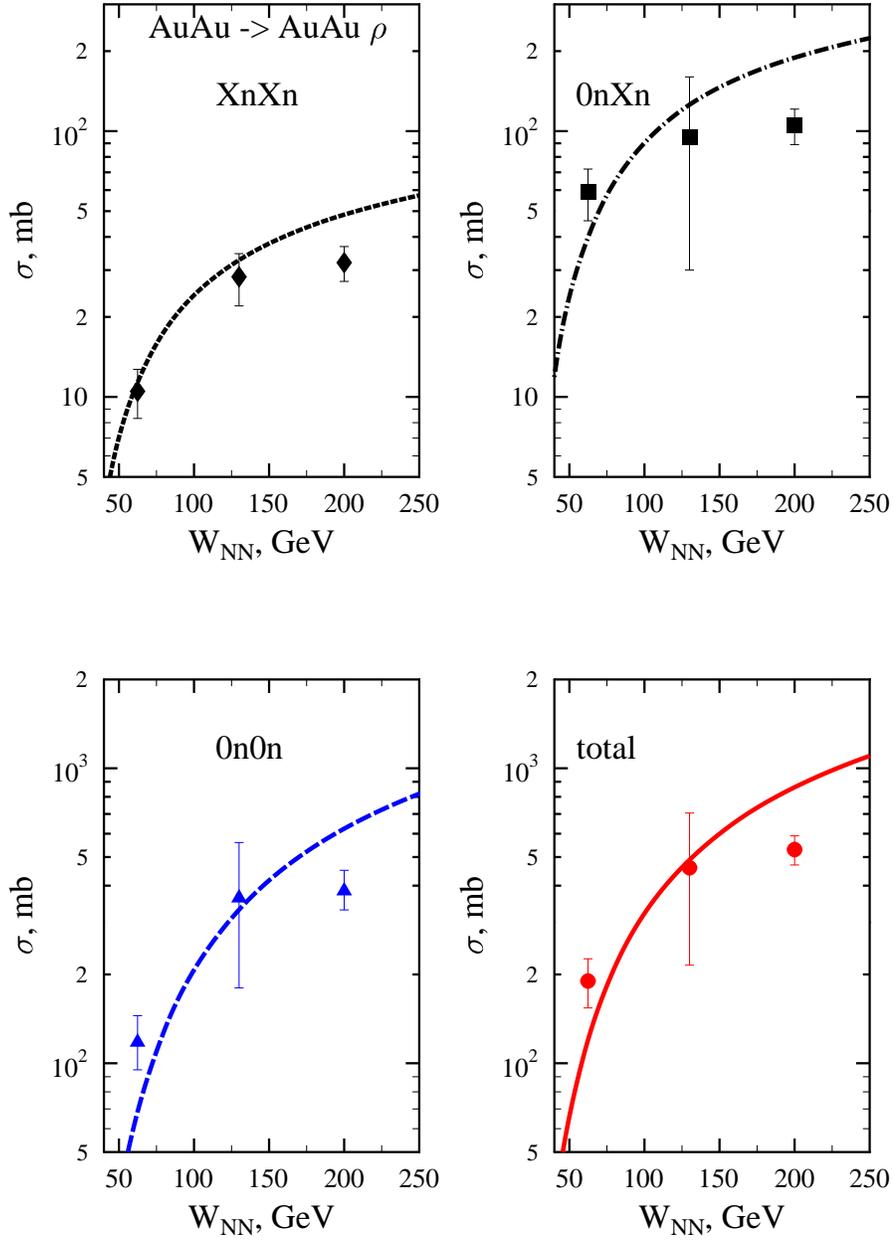, height=18cm}
 \caption{Comparison of the calculated cross sections for $\rho$ photoproduction in the
gold-gold UPC with the STAR experimental results.}
 \label{rothexp}
\end{center}
\end{figure}

\begin{figure}
\begin{center}
\epsfig{file=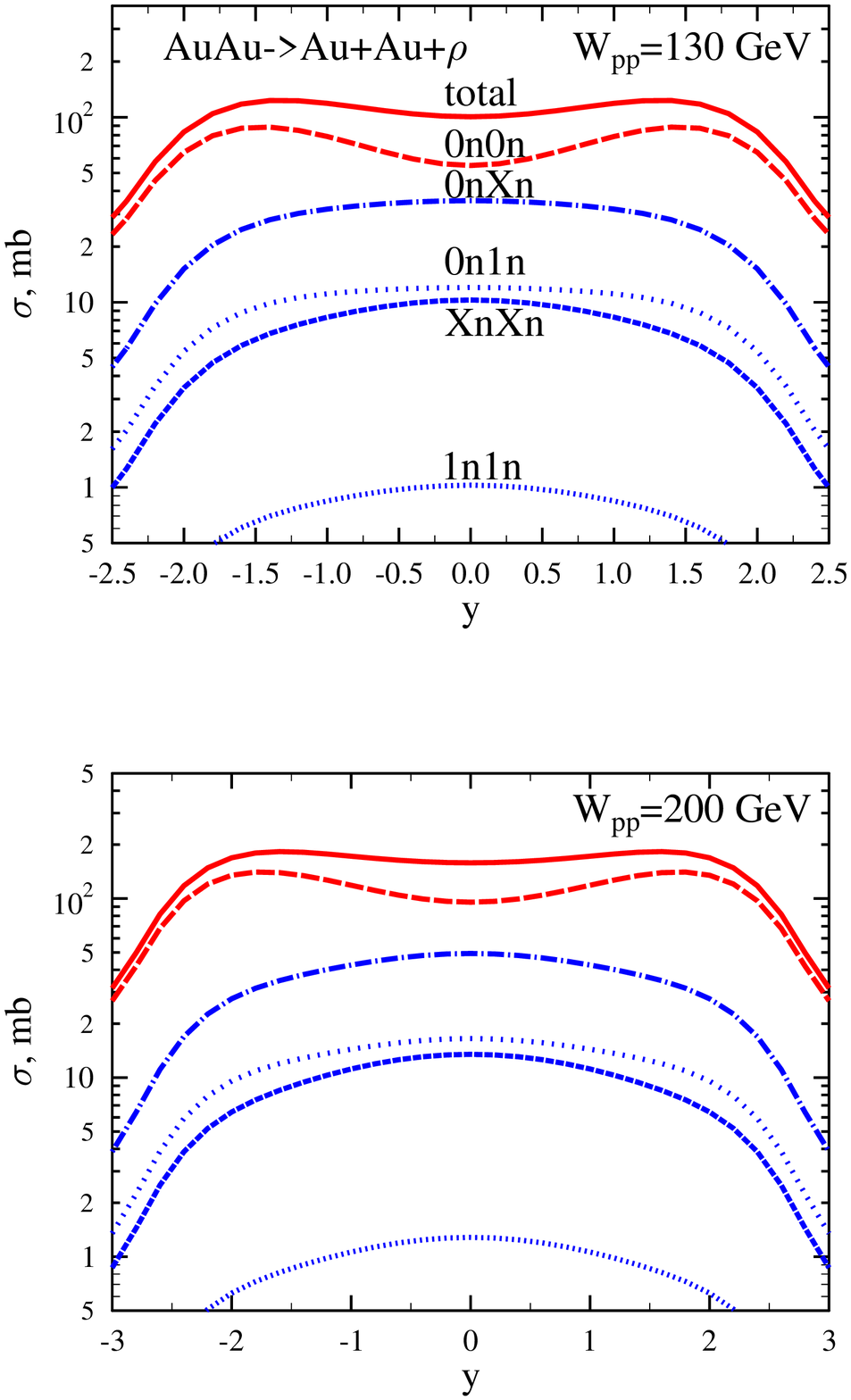, height=18cm}
 \caption{Calculated rapidity distributions for different channels in $\rho$ photoproduction
in gold-gold UPC at RHIC.}  
 \label{rorhic}
\end{center}
\end{figure}

\begin{figure}
\begin{center}
\epsfig{file=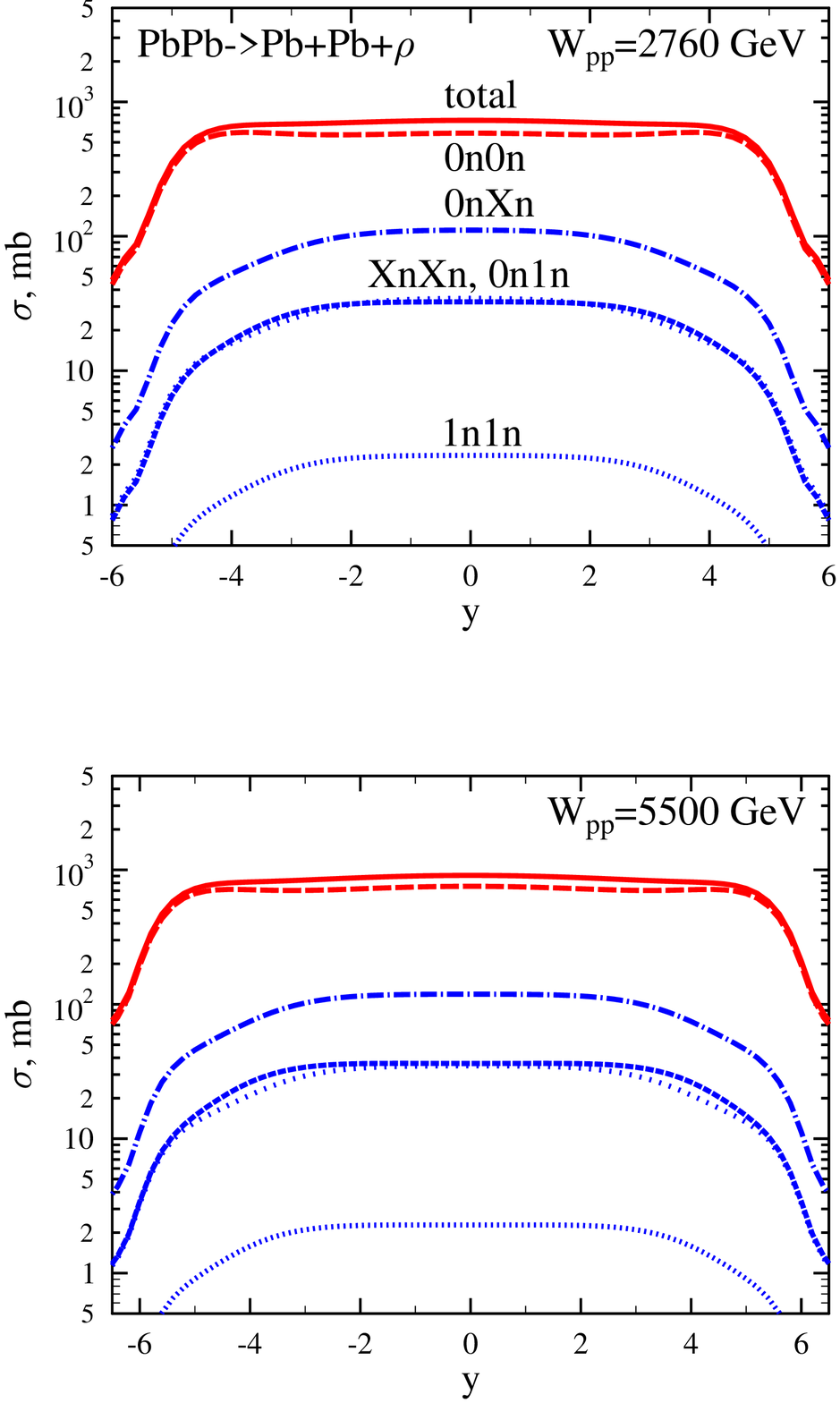, height=18cm}
 \caption{Calculated rapidity distributions for different channels in $\rho$ photoproduction
in lead-lead UPC at LHC.} 
 \label{rolhc}
\end{center}
\end{figure}

Production of $\rho$-meson provides an important check of the modeling of the dynamics of UPC. 
Indeed, the cross section $\gamma +A \to \rho + A$ 
is well understood theoretically. Though the Gribov space -time picture of high energy process of 
hadron - nucleus interaction differs  qualitatively from low energy Glauber picture,  
numerically,  the Gribov-Glauber theory which includes inelastic screening effects  
predicts very small (on the level of few percent) deviations from the Glauber formula 
for  hadron-nucleus scattering\footnote {For the recent discussion see \cite{Alvioli:2008rw}}.
An additional element of the theory in the case of the photon projectile  
is presence of the non-diagonal 
transitions $\gamma \to \rho' \to \rho$.  The 
generalized vector dominance model combined with Glauber approach describes well the data 
on $\rho$ production at $E_\gamma \sim 10 GeV$ without free parameters
\cite{Frankfurt:2002wc}.
It also predicted \cite{Frankfurt:2002sv} reasonably well absolute cross section 
of the $\rho$ meson photoproduction for $\sqrt{s_{NN}}$ =130 GeV as measured 
in the UPC at RHIC \cite{Adler:2002sc}. For high energy photoproduction off heavy 
nuclei correction to standard VDM+Glauber model due to the account for the 
nondiagonal transitions is about $10\div 15\%$. Since
 the uncertainty (combined experimental errors due to the restricted statistics and systematics) 
of the experimental cross sections measured by STAR
exceeds this number almost by factor two,
we use the standard Glauber model formula \cite{Bauer:1977iq} for calculation of the cross section 
of coherent $\rho$ meson photoproduction in the kinematics of UPC of heavy ions at RHIC and LHC
:
\begin{equation}
\sigma_{\gamma A\to \rho A}=\frac {d\sigma_{\gamma N\to \rho N}(t=0)} {dt}
\int \limits_{-\infty}^{t_{min}} dt \biggl |
\int \limits_{0}^{\infty} d\,{\vec b} e^{i{\vec q}_{\bot }\cdot {\vec b}}
\int \limits_{-\infty}^{\infty}dz \,\varrho ({\vec b},z)
e^{iq_{\| }^{\gamma \to \rho}z}e^{-\frac {\sigma_{\rho N}} {2}
\int \limits_{z}^{\infty} \varrho ({\vec b},z')dz'}\biggr |^2.
\label{gampl} 
\end{equation}
Here 
${\vec q}_{\bot }^2=t_{\bot }={t_{min}-t}$, 
$-t_{min}=|q_{\| }^{\gamma \to \rho}|^2=\frac {M_{\rho }^4} {4q_0^2}$ 
is longitudinal momentum transfer in $\gamma -\rho $ transition.
We also use Donnachie-Landshoff model \cite{Donnachie:1999yb}  
to calculate the total $\rho N$ cross section  
and forward cross section of coherent $\rho$ photoproduction off 
the nucleon target. 
This model describes reasonably well the data for $\gamma p\to p \rho$ at energies
up to $W_{\gamma p} \approx 200\div 300$ GeV studied at HERA but can 
slightly overestimate the
cross section at the very high energies due to introducing the hard pomeron exchange with
large intercept ($\alpha_{H}=1.44$)
in such soft process as coherent photoproduction of $\rho$ with small transverse momentum. 
Note, that we neglect the interesting effect of interference between production by 
left and right moving photons \cite{Klein:1999gv} since it gives a small correction 
in the cross section integrated over 
rapidities, $p_t$.
Our calculations for $\rho$ meson photoproduction in UPC 
at RHIC and LHC are given in Tables \ref{tcrsec}-\ref{tcrseclhc} 
and Figures \ref{rothexp}-\ref{rolhc}.
In Table \ref{tcrsec}  and Fig. \ref{rothexp} we compare cross sections obtained in 
the model described above
with results presented by the STAR collaboration.

It is worth noting
that in the  STAR experiment only the  (XnXn) and (1n1n) channels
were measured. The $\rho$-meson  were detected in the 
rapidity range $-1<y<1$ and $p_t <150$ MeV/c. The numbers reported by experiment 
for the total cross sections of various channels were calculated from the measured 
two cross sections using the the StarLight event generator to extrapolate to 
the larger $|y|$ and to calculate the ratio of the cross section in different channels.

  We find a reasonable agreement at lower energies but a significant discrepancy in all 
measured partial cross sections at $\sqrt{s_{NN}}=200 $ GeV. 

 \begin{table}[here]
    \begin{center}
      \begin{tabular}{|c|c|c|c|}\hline
	$\sqrt s_{NN}$&  62.4 GeV & $130 \,GeV$& $200\, GeV$\\ 
           UPC             & AuAU  & AuAu & AuAu  \\ \hline
	$\sigma_{coherent}^{\rho}$ , mb  & 137  &520 &910        \\
        STAR                 &(190$\pm $36)& (460$\pm $245)&(530$\pm $60)\\\hline
	$\sigma_{0n0n}^{\rho}$ , mb  &79  & 354 &661  \\ 
        STAR                 &(120 $\pm 25 $)&(370$\pm 90  $)&(39$\pm 60$) \\\hline
	$\sigma_{0nXn}^{\rho}$ , mb  &45  & 132 &198  \\ 
        STAR                  &(59.3 $\pm 13 $)&(95$\pm 65  $)&(105$\pm 16$) \\\hline
	$\sigma_{XnXn}^{\rho}$ , mb  &13  & 34 &51  \\ 
        STAR                  &(10.5 $\pm 2.2 $)&(28.3$\pm 6.3  $)&(32$\pm 4.8$) \\\hline
	$\sigma_{0n1n}^{\rho}$ , mb  &16 & 46&67 \\ 
        STAR                  & - & - & - \\\hline
	$\sigma_{1n1n}^{\rho}$ , mb  &1.4 & 3.5 &4.9  \\
        STAR                  & - & (2.8$\pm 0.9  $)& (2.4$\pm 0.5$) \\\hline
      \end{tabular}
    \end{center}
    \caption{Cross sections of the
$\rho$ meson photoproduction in gold-gold UPC at RHIC 
calculated in the Glauber model. Numbers in brackets present results obtained by STAR collaboration
from analysis of the experimental data on the gold-gold UPC studies at RHIC.}
    \label{tcrsec}
  \end{table}

 \begin{table}
    \begin{center}
      \begin{tabular}{|c|c|c|c|} \hline
energy	&Partial cross section, mb&  $\sigma (XnXn, |y|<1)$ & $\sigma ( 1n1n,|y|<1)$ \\ \hline
 $\sqrt s_{NN} =130  $      &Glauber model  &   18.6    & 1.87       \\ 
  GeV	     &STAR experiment  & 14.9 $\pm 2.0  $ & 1.47 $\pm 0.16  $ \\ \hline
$\sqrt s_{NN} =200$	&Glauber model  &   25.4     & 2.4       \\ 
GeV	&STAR experiment  & 14.5 $\pm 2.0  $ & 1.07 $\pm 0.16  $ \\ \hline
      \end{tabular}
    \end{center}
    \caption{Comparison of the Glauber model predictions with the results of measurements by STAR for
$\rho$ meson photoproduction in gold-gold UPC at $\sqrt s_{NN} =200 $ GeV at RHIC.}
    \label{tcrsec1}
  \end{table}

In particular, the data reported  \cite{Abelev:2007nb} do not  
show an increase of the cross section with 
increase of energy from $\sqrt s_{NN} =130$ GeV to $\sqrt s_{NN} =200$ GeV expected 
in {\bf all} theoretical calculations. 
This is puzzling since the energy 
dependence of the cross section of $\rho$ photoproduction in UPC of heavy nuclei 
comes practically solely from the increase of the photon flux 
which is essentially model independent and,
 hence,  very similar in all calculations.  
 The cross section $\gamma +p \to \rho +p$ is a weak function of energy in the discussed energy 
interval and $\gamma +A \to \rho +A$ is expected to be 
even weaker  function of the incident energy due to blackness of the interaction 
with  heavy nuclei at small impact parameters.

 One can see from the Table \ref{tcrsec1} that 
the measured cross sections at two energies are practically equal while 
the calculated  ones  increase 
by a factor $\approx 1.3$, which  coincides with an increase of the average photon  flux.
 
 In Fig.\ref{rolhc} we present our predictions for the rapidity distributions 
of the $\rho$ photoproduction at energies $\sqrt s_{NN} =2.76$ TeV and $\sqrt s_{NN} =5.5$ TeV at LHC.  
The total cross sections are given in Table \ref{tcrseclhc}.

 \begin{table}
    \begin{center}
      \begin{tabular}{|c|c|c|}\hline
	$\sqrt s_{NN}$& $2760\, GeV$& 5500 GeV \\ 
           UPC             & PbPb  &PbPb   \\ \hline
	$\sigma_{coherent}^{\rho}$, mb  &7023& 9706       \\ \hline
	$\sigma_{0n0n}^{\rho}$ , mb  &5915&8309 \\ \hline 
	$\sigma_{0nXn}^{\rho}$ , mb  &847&1057 \\ \hline
	$\sigma_{XnXn}^{\rho}$ , mb  &261&340 \\ \hline 
	$\sigma_{0n1n}^{\rho}$ , mb  &260&306 \\ \hline 
	$\sigma_{1n1n}^{\rho}$ , mb  &18.5&21 \\ \hline
      \end{tabular}
    \end{center}
    \caption{Cross sections of the
$\rho$ meson photoproduction in  PbPb UPC at LHC
calculated in the Glauber model. }
    \label{tcrseclhc}
  \end{table}

Since $d\sigma(\gamma A\to \rho A )/dt(t=0)$  weakly depends on energy, 
combined studies at several energies and at different rapidities will provide a stringent test of 
the dynamics of the break up of nuclei due to e.m. interactions.
 
 Note in passing that our cross sections larger than the Starlight results since the Starlight 
MC code \cite{starlight}uses expression 
for the total cross section of hadron - nucleus interaction given by classical mechanics which in the 
limit of large total $VN$ cross sections and large A differs from the Gribov-Glauber model 
by a factor of  about two. Recently one more update of predictions
for total coherent cross sections in heavy ion UPC at RHIC and LHC
has been published \cite{Goncalves:2011vf}. The calculations in \cite{Goncalves:2011vf}
are based on use of Color Glass Condensate (CGC) ideas and two versions of 
the color dipole model (IP-SAT \cite{ipsat}
and IIM\cite{iim}).  Comparing to the data of STAR at $\sqrt s_{NN}=200$ GeV the authors
found that IP-SAT model gives a larger cross section of $\rho$ photoproduction but 
the result with IIM model appears to be close to that of the STAR experiment. So, cross section at the LHC energies 
predicted with this model are considered by the authors as preferable. However, as we show
in this paper, the STAR measurements at $\sqrt s_{NN}=200$ GeV give the cross sections
practically equal to cross sections at $\sqrt s_{NN}=130$ GeV, and this doesn't agree with reasonable
energy dependence, dictated by the increase of the photon flux with increase of $\sqrt s_{NN}$.
As a result, the cross sections of $\rho$ photoproduction in UPC at LHC, 
predicted in \cite{Goncalves:2011vf}, appeared to be smaller than our. 
Note in passing that the color dipole models are usually used for description 
of the hard interactions. A rational to apply such a model for the description 
of the soft process of the $\rho$ meson exclusive photoproduction is not clear,
as opposed to the use of the (generalized) vector dominance model which is 
theoretically well justified in this case.

\begin{figure}
\begin{center}
\epsfig{file=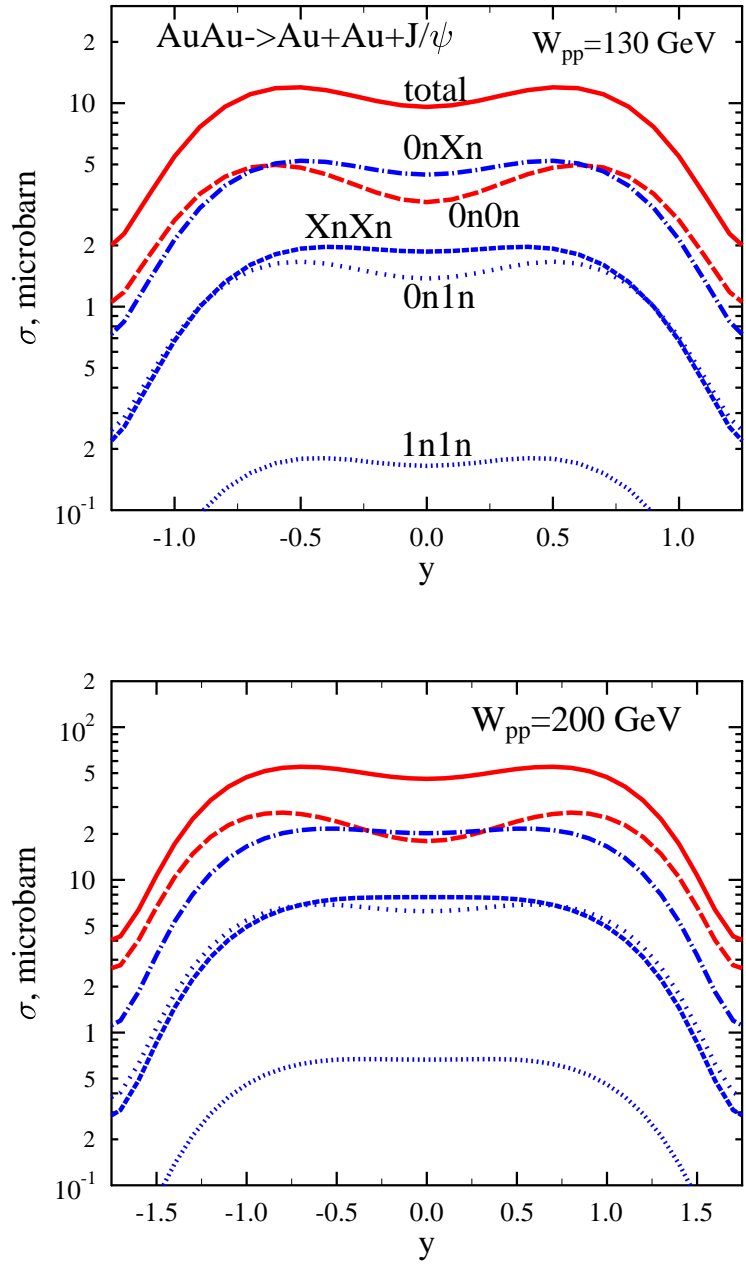, height=18cm}
 \caption{Calculated rapidity distributions for different channels in $J/\psi$ photoproduction
in gold-gold UPC at RHIC.}
 \label{psirhic}
\end{center}
\end{figure}

\begin{figure}
\begin{center}
\epsfig{file=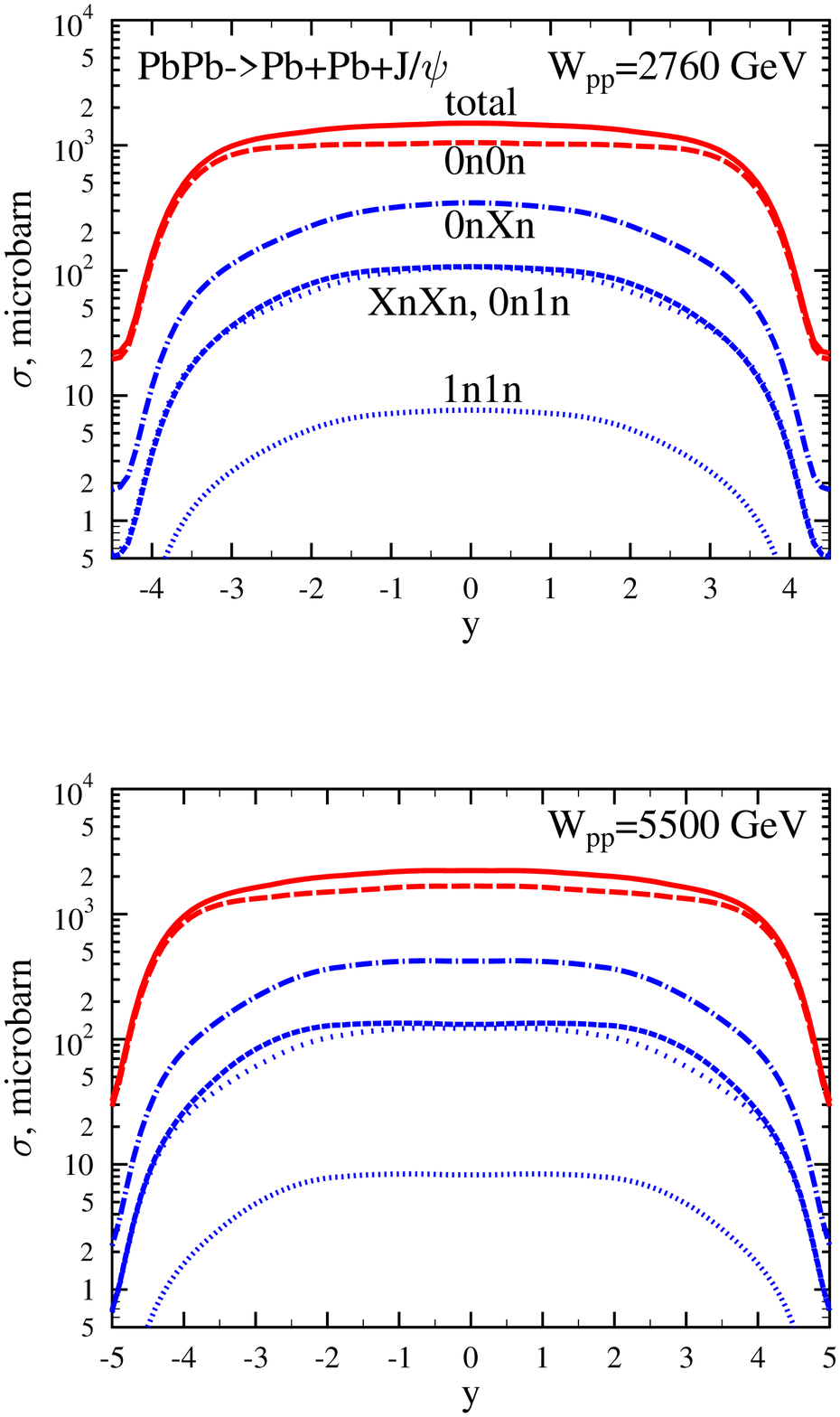, height=18cm}
 \caption{Calculated rapidity distributions for different channels in $J/\psi$ photoproduction
in lead-lead UPC at LHC.} 
 \label{psilhc}
\end{center}
\end{figure}

\begin{figure}
\begin{center}
\epsfig{file=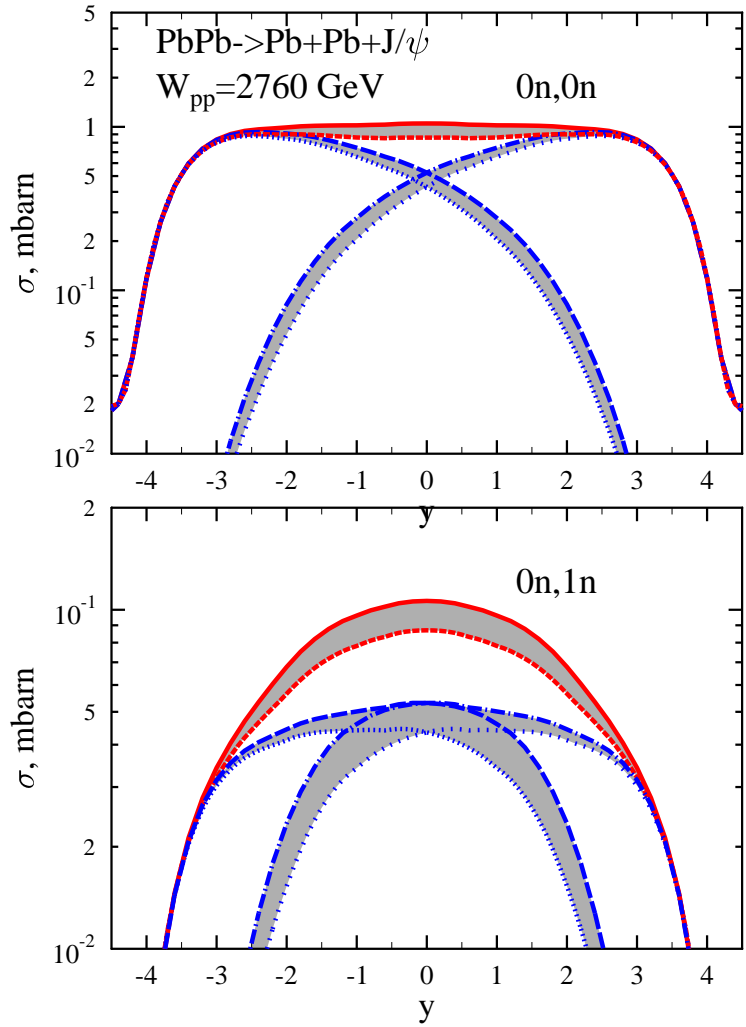, height=18cm}
 \caption{Calculated rapidity distributions for (0n0n) and (0n1n) channels in $J/\psi$ photoproduction
in PbPb UPC at $\sqrt s_{NN}=2.76$ TeV. Shaded area shows uncertainty in accounting for
the gluon nuclear shadowing. Left and right curves demonstrate one-side contributions.}
 \label{psi01}
\end{center}
\end{figure}

\begin{figure}
\begin{center}
\epsfig{file=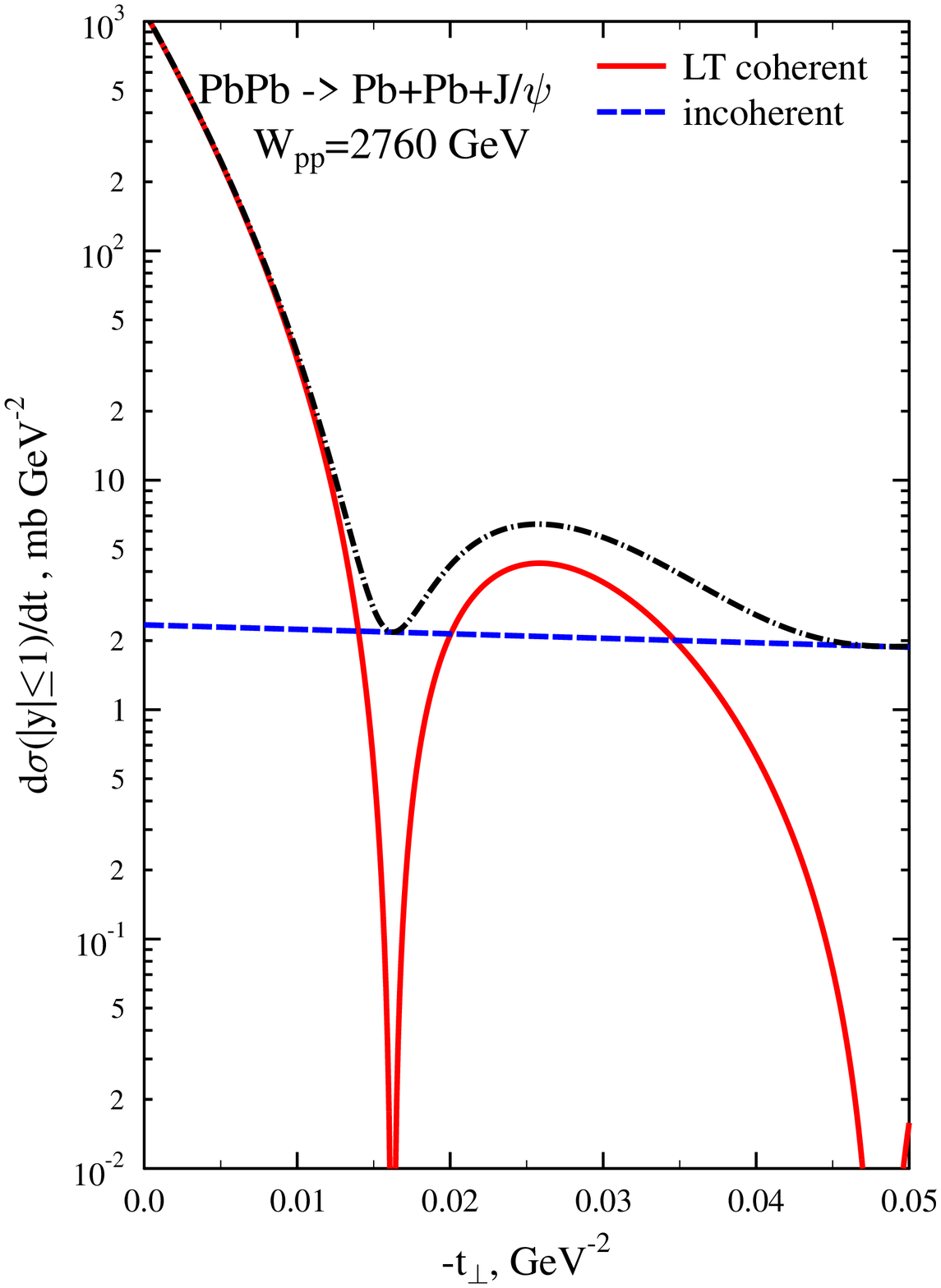, height=18cm}
 \caption{Distributions on the transverse momentum transfer in coherent and incoherent
$J/\psi$ photoproduction in PbPB UPC at $\sqrt s_{NN}=2.76$ TeV.}
 \label{psidt}
\end{center}
\end{figure}

\section{$J/\psi$ production}

The coherent production of $J/\psi$ in UPC is one of the direct ways of probing the pattern of the 
interaction of small color dipoles of average 
transverse size $\sim 0.2 \div 0.25 fm$ \cite{Frankfurt:2008et}. Due to their 
small size, the dominant mechanism of coupling of the $c\bar c$ to the nucleon is through a 
two gluon attachment which forms a start of the gluon ladder. In  
the dipole approximation the cross 
section of $\gamma + T\to J/\psi +T$ process can be calculated in the Leading Order of PQCD
\cite{Ryskin:1992ui,Brodsky:1994kf}
\begin{equation}
\frac {d\sigma_{\gamma T\rightarrow J/\psi T}(t=0)} {dt} =\frac {16\Gamma_{ee}{\pi}^3} {3{\alpha_{em}}M_{J/\psi}^5}
\biggl [\alpha_{s} (\mu^2 )xG_T (x,\mu^2 )\biggr ]^2
\label{dcslo}
\end{equation} 
Here $\Gamma_{ee}$ is width of the leptonic decay of $J/\psi$ and $G_T (x,\mu^2  )$
is density of gluons with fraction $x=\frac {M_{J/\psi}^2} {s_{\gamma T}}$ of momentum of the target
($s_{\gamma T}$
is the invariant energy for $\gamma - T$
scattering), and $\mu^2$ is the scale which we choose to be $\mu^2 =\frac {M_{J/\psi}^2} {4}$.
Applying Eq.\ref{dcslo} to the proton and nucleus target, 
one can write the cross section for the photoproduction
of $J/\psi$ off the nucleus in the form 
\begin{equation}
{\sigma_{\gamma A\rightarrow V A}(\omega)=  
{d \sigma_{\gamma N \to V N}(\omega,t_{min})\over dt}
\Biggl [\frac {G_{A}(\frac {M_{V}^2} {s},Q^2)}
{G_{N}(\frac {M_{V}^2} {s},Q^2)}\Biggr ]^2
\int \limits_{-\infty}^{t_{min}} dt
{\left|
\int d^2bdz e^{i{\vec q_{\perp}}\cdot {\vec b}}
e^{-q_{\parallel}z}\rho ({\vec b},z)
\right|
}^2}.
\label{phocs}
\end{equation}

The key feature of the photoproduction
of the hidden heavy flavor vector mesons off heavy nuclei is the gluon nuclear shadowing which
is characterized by the ratio of the nuclear gluon density distribution $G_{A}(x,\mu^2)$ to the proton
gluon distribution $G_{N}(x,\mu^2)$. 

The $G_A/G_N$ ratio can be calculated within the theory of leading twist 
nuclear shadowing \cite{FS99}( see \cite{Frankfurt:2011cs} for details of the model and references )
developed on the base of the Gribov's theory of inelastic shadowing, Collins's factorization
theorem for hard diffraction and the DGLAP evolution equations.
 This model uses, as an input,
nucleon diffractive parton distribution functions (pdf) which are available both in the NLO and in LO.  
In our study we select LO nucleon pdfs which give a reasonable description of the energy 
dependence of the elementary cross section. Also, we use the LO gluon density in proton
$G_N (x,\mu =2.5 GeV^2)$,
recently found \cite{Martin:2007sb} by the Durham-PNPI group in the 
range $10^{-4}<x<10^{-2}$ from the fit of cross section, given by
(Eq.\ref{dcslo}), to the HERA experimental data on $\gamma +p\rightarrow J/\psi +p$.

The results of calculations are presented in Figs. \ref{psirhic} - \ref{psi01} and in the Table \ref{tcrspsi}.
Since the estimate of shadowing is based on the parton distribution functions determined from
the data, there is some uncertainty related to the experimental errors. In Fig. \ref{psi01} we show how
this uncertainty influence our estimates of the cross sections.

Comparing our results with those of StarLight and \cite{Goncalves:2011vf}, we find for
$J/\psi$ photoproduction at LHC smaller cross sections. This is due to the accounting 
for the Leading Twist gluon shadowing which was not included in  StarLight.  
The calculation of \cite{Goncalves:2011vf}
included only  higher twist shadowing in the color dipole  model which is known 
to be much smaller than the leading twist shadowing in a wide range of $x$, 
see \cite{Frankfurt:2002kd} for detailed comparison of the shadowing in 
the leading twist approximation and in the color dipole model.

As we mentioned before, the coherent scattering is determined by  
the sum of lower and higher energy contributions.
 This makes it difficult to separate these contributions at any rapidity, 
  except for $y=0$ where they are equal.   
  One would like to determine the coherent cross section in as wide range
   of $\gamma A$  energies as possible, with higher energies allowing to probe 
 the gluon density at lower $x$. 
 For example,
in PbPb UPC at $\sqrt s_{NN}=2.76$ TeV and rapidity of produced $J/\psi$ $y\approx 0$ the
corresponding $x\approx 10^{-3}$, while, if one could extract contribution due to
the higher energy photon to the 
 the cross section of production $J/\psi$ with
$y\approx 2$   , it would be possible to  extract gluon density at $x\approx 1.5\cdot 10^{-4}$.
However, due to the rapid drop of the flux of photons with higher energy, this contribution is
significantly lower than the cross section from the low energy photons (see Fig.\ref{psi01}) which is determined
by the gluon density at $x\approx 10^{-2}$. It was argued\cite{Baltz} that measurement of the process 
of photoproduction with
additional photon exchange enhances the contribution from the higher energy photons.
We would like to emphasize that measuring at the same $\sqrt s_{NN}$ two or more channels with
$J/\psi$ at fixed rapidity,
for example, (0n1n) and (0n0n) or any other, one can easily separate
 high and low energy
  contributions, provided that
the fluxes of photons modified by accounting for the neutron decay are calculated with
reasonable accuracy. These fluxes and the procedure in whole can be tested in study of $\rho$
photoproduction in PbPb UPC since the cross section of $\rho$ photoproduction off nuclear target
weakly depends on energy and can be calculated in the Glauber model. Also we would like to note
that contribution of the incoherent photoproduction in the UPC at LHC which are usually
accompanied by neutrons \cite{Strikman:2005ze}  to the coherent cross section with
detection of neutrons will be very small if one uses the low $p_t$ of produced $J/\psi$ 
(see Fig. \ref{psidt}).

 \begin{table}
    \begin{center}
      \begin{tabular}{|c|c|c|c|c|}\hline
	$\sqrt s_{NN}$  &AuAu $130 GeV$&AuAu $200 GeV$&PbPb $2760 GeV$&PbPb 5500 GeV \\ \hline
	$\sigma_{tot},\, \mu b$  &24&137&9281& 15836      \\ \hline
	$\sigma_{0n0n}, \, \mu b$ & 10.1 & 68 & 7044 & 12376 \\ \hline
	$\sigma_{0nXn}, \, \mu b$ & 10.2 & 52 & 1689 & 2584 \\ \hline
	$\sigma_{XnXn}, \, \mu b$ & 3.7 & 17 & 548 & 876 \\ \hline
	$\sigma_{0n1n}, \, \mu b$ & 3.25 & 16.8 & 512 & 739 \\ \hline
	$\sigma_{1n1n}, \, \mu b$ & 0.35 & 1.6 & 39 & 54 \\ \hline
      \end{tabular}
    \end{center}
    \caption{Cross sections of
$J\psi$ photoproduction in gold-gold UPC at RHIC and PbPb UPC at LHC
calculated in the leading order LT shadowing model.}
    \label{tcrspsi}
  \end{table}

 If the measurements are done at two energies
$\sqrt{s}\mbox{= 2.76 TeV}$
and 
 $\sqrt{s}\mbox{= 5.52 TeV}$  
 a simultaneous analysis of two data sets will allow to extend 
  model independent determination of nuclear 
ratio  to $x$ at least a factor of two smaller than  in $y=0$ at 
$\sqrt{s}\mbox{= 5.52 TeV}$.

\section{Conclusions}

The studies of the $J/\psi$ production in the heavy ion UPC 
at the LHC will allow to measure coherent 
photoproduction cross section down to $x\approx  10^{-4}$ using the channels without 
nucleus break up which dominate at the LHC energies. A cross check will be possible 
using channels with dissociation of nuclei. However, this would require sorting out 
situation with the $\rho$-meson production at the RHIC energies. 
This requires both further experimental studies   at  RHIC and LHC as 
well as further theoretical studies.

We would like to thank L.Frankfurt, V. Guzey, S.Klein, J.Nystrand, Yu.Gorbunov,
for the discussions.
This work was supported in part by the US DOE Contract Number
DE-FG02-93ER40771 and by Program of Fundamental Researches at LHC
of Russian Academy of Science.

\end{document}